\documentclass[sigconf,nonacm]{acmart}
\usepackage{amsmath,amsfonts}
\usepackage{algorithmic}
\usepackage{graphicx}
\usepackage{textcomp}
\usepackage[table]{xcolor}
\usepackage{orcidlink}
\usepackage{float}
\usepackage{subcaption}
\usepackage{tabularx}
\newcolumntype{Y}{>{\centering\arraybackslash}X}
\usepackage{tikz}
\usetikzlibrary{arrows.meta, positioning, shapes.geometric}
\usepackage{afterpage}
\usepackage{booktabs} 
\usepackage{microtype} 
\usepackage{enumitem}
\usepackage{siunitx}
\usepackage{xspace} 
\usepackage{soul} 
\usepackage{adjustbox} 
\usepackage{array,multirow}
\usepackage{colortbl}
\usepackage{threeparttable} 
\usepackage{tablefootnote}
\usepackage{comment}
\usepackage{wrapfig}
\usetikzlibrary{patterns}
\usetikzlibrary{shapes,arrows}
\usetikzlibrary{positioning,calc}
\usetikzlibrary{decorations.pathreplacing}
\usetikzlibrary{shapes.geometric,shapes.arrows,decorations.pathmorphing}
\usetikzlibrary{matrix,chains,scopes,positioning,arrows,fit,backgrounds}
\usepackage{longtable}
\usepackage[utf8]{inputenc}
\usepackage{tcolorbox}
\usepackage{listings} 
\usepackage{makecell}
\usepackage[ruled]{algorithm2e}
\usepackage{pifont}
\usepackage[version=4]{mhchem} 
\usepackage[labelfont=bf,textfont=normalfont]{caption}
\usepackage{cleveref}

\newcommand{\ignore}[1]{}

\definecolor{codegreen}{rgb}{0,0.6,0}
\definecolor{codegray}{rgb}{0.5,0.5,0.5}
\definecolor{codepurple}{rgb}{0.58,0,0.82}
\definecolor{backcolour}{rgb}{0.95,0.95,0.92}

\lstdefinestyle{mystyle}{
  backgroundcolor=\color{backcolour}, commentstyle=\color{codegreen},
  keywordstyle=\color{magenta},
  numberstyle=\tiny\color{codegray},
  stringstyle=\color{codepurple},
  basicstyle=\ttfamily\footnotesize,
  breakatwhitespace=false,         
  breaklines=true,                 
  captionpos=b,                    
  keepspaces=true,                 
  numbers=left,                    
  numbersep=5pt,                  
  showspaces=false,                
  showstringspaces=false,
  showtabs=false,                  
  tabsize=2
}

\tikzset{every node/.style={font=\small}}
\AtBeginDocument{%
  }

\begin{document}

\title{Towards Low-Cost Low-Power Activity-Aware Soil Moisture Sensing Platform for Large-scale Farming}

\author{%
  Jack Thoene\orcidlink{0009-0001-8719-2126},
  Omar Kamil,
  Thekra Alkadee,
  Nivedita Arora\orcidlink{0000-0002-8630-9919}%
}
\affiliation{%
  \institution{Embodied System Lab,  McCormick School of Engineering, Northwestern University}%
  \city{Evanston}%
  \state{Illinois}%
  \country{USA}%
}

\renewcommand{\shortauthors}{Thoene et al.}

\begin{abstract}
Deep understanding of a field's soil moisture content is the leading indicator for predicting crop yields and making data driven decisions for irrigation and application of topical chemicals for drought resilience. Despite this importance, the cost of adopting and maintaining IoT infrastructure prevents modern farms from employing widespread real time soil moisture sensors. We present an end-to-end platform of buried battery-free sensor nodes and a mobile basestation that leverages the farmer’s daily routine for data retrieval. Each node features a self-powered galvanic soil-moisture probe, employing a high impedance analog front end to enable durability.  Operating entirely on harvested solar energy for up to 21 days on a single capacitor charge, each node collects soil moisture, temperature, and environment condition data. Using a predictable finite-state machine, handshake-based data exchanges occur with a basestation affixed to standard farming vehicles designed to listen for the nodes while moving through the farm. Our platform organizes all sensor, link-quality, and location data into an easy-to-interpret dashboard to seamlessly integrate with the farmer's everyday routine. Costing less than \$35, the platform is a financially accessible, accurate, and easily scalable platform that enables persistent, regular data collection from the most rural plots without adding to or impeding farming operations. Experimental evaluation demonstrates reliable communication over 1 km at 2 dBm transmit power, stable sensor readings over 70 days of indoor operation, and continuous data recovery during multiple periods of intermittent connection. 

\end{abstract}




\maketitle 

\begin{figure}[!t]
  \centering
  \includegraphics[width=\linewidth]{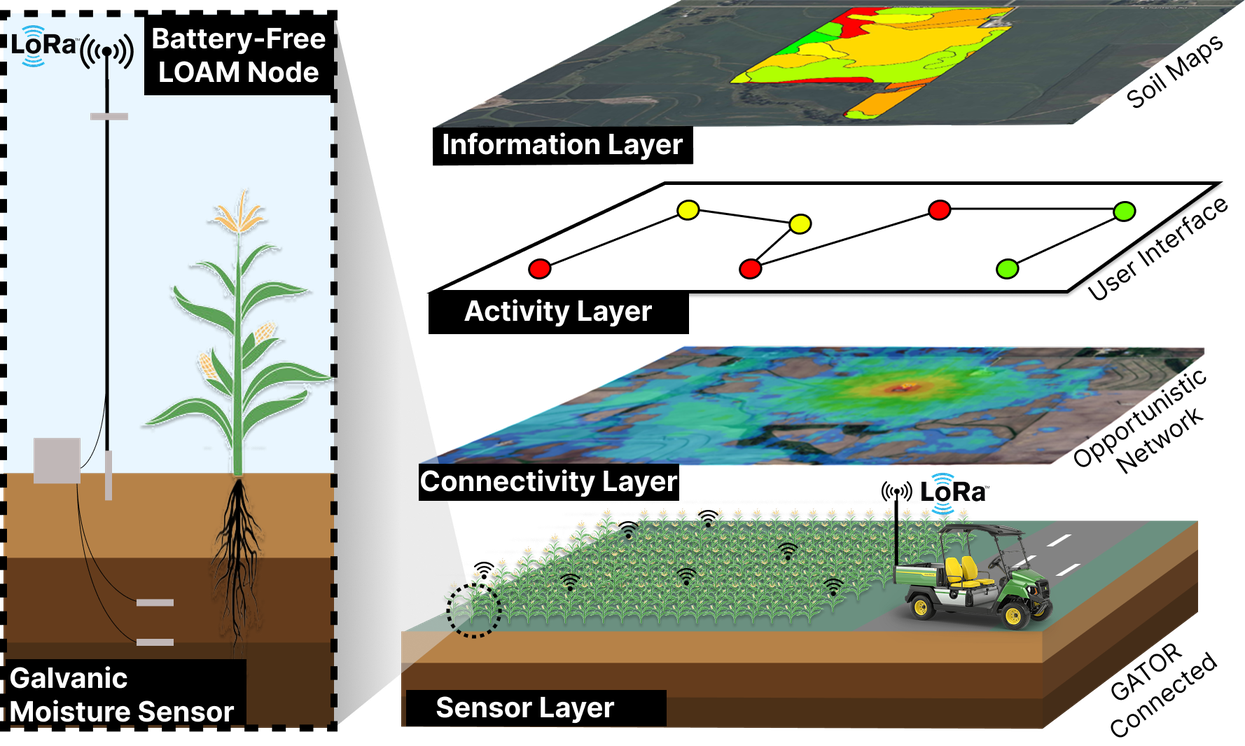}
  \caption{\textbf{Our platform} is a four-layer framework designed to support distributed soil moisture sensing across large farms.
At the \textit{sensor layer}, LoRa-based, battery-free soil moisture sensors manage their communication, sensing, and energy harvesting schedules.
The \textit{connectivity layer} leverages the opportunistic movement of the farmer’s Gator vehicle, which acts as a mobile base station or “data mule” to collect sensor data.
During routine farm inspections, the \textit{activity layer} allows the farmer to plan their path to opportunistically gather soil moisture readings across the field.
Finally, the \textit{information layer} processes this data to alert the farmer when moisture levels are low and can generate soil moisture maps to guide long-term decision-making.
}
\label{fig:firstfigure}
\end{figure}

\section{Introduction}
The United Nations Sustainable Development Goal 2, Zero Hunger, calls for increasing food production to feed a growing global population while preserving critical resources \cite{un_sdg2_zero_hunger}. Soil moisture has long served as a primary metric for soil and crop health. More recently, it has become central to improving crop yield \cite{holzman_estimating_2014} by informing when to irrigate \cite{moody_nonlinear_1966,duffkova_effect_2022, soulis_investigating_2015, stafford_soil_2013} and when to apply topical sprays that temporarily improve crop resilience to heat stress during drought \cite{hu_effect_2008}. Monitoring changing soil conditions with wired sensors and gateways in real time is straightforward where power and connectivity are readily available, such as in seeded research plots, greenhouses, canalized irrigation systems, and drip-line installations. However, \textbf{large-scale commercial farms operate under extreme constraints of power and wireless connectivity}: they often extend for kilometers without permanent roads, power lines, cell coverage, or irrigation piping \cite{TRIP_2020_RuralConnections}, making isolated rural plots difficult to monitor. 

To address this isolation, many modern farms rely on remote sensing tools like satellite imaging \cite{yang2018high}. While these offer high-resolution visuals, they are costly and limited to single snapshots, often reserved for events like storm damage assessment. Some tractors integrate GPS and soil probes to map soil variation \cite{noauthor_sensing_nodate}, but these are only used during planting or harvest, a brief 7-day window in the farming cycle. After planting, farmers rarely revisit field interiors until harvest, making continuous soil monitoring impractical. As a result,  most farmers even today rely on frequent visual inspection of roadway-accessible edges in their ruggedized golf cart, leaving vast interior regions completely unmonitored. 

Wireless IoT solutions present the possibility of overcoming these issues and provide \textbf{regular datastreams from even the most remote areas of the farm.} Cellular based networks have been shown to be viable for km scale urban monitoring \cite{daepp2022eclipse}, but they  exhibit significant coverage lapses in rural regions and densely vegetated inner farmlands, even in high income countries \cite{usda_broadband, pender2023three, fcc_rural_call_completion}. Coverage is typically optimized along highways and power corridors rather than over cropland. Extending service requires new tower infrastructure whose \textbf{installation and data ownership} are controlled by external providers rather than the farmer. 

\vspace{-0.1in}
\begin{figure}[h]
    \centering
    \includegraphics[width=\linewidth]{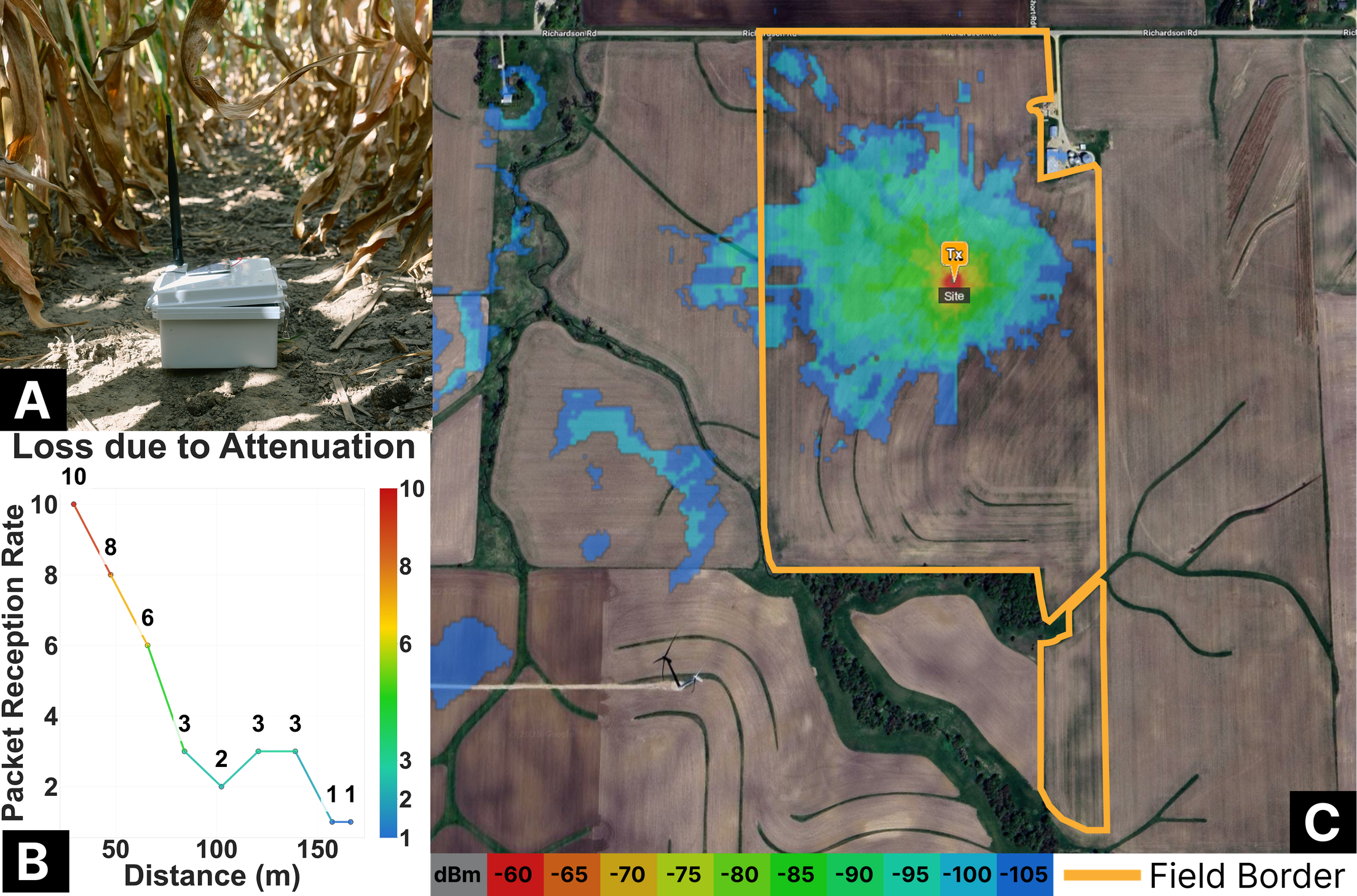}
    \caption{\textbf{Effect of vegetation obstruction on LoRa wireless range in a real corn field} {(A)} Deployment of a low-power (2 dBm) LoRa node at ground level 1 month before harvest. {(B)} Packets received vs. distance limits transmission inside of 250 m. (C) Transmission coverage across the field from a single node.} 
    \label{fig:figure2}
\end{figure}
\vspace{-0.1in}

To overcome these limitations, Low Power Wide Area Networks (LPWANs) like LoRa \cite{bor2016lora} offer a way to extend network range from a small number of gateways to many sensor nodes distributed across the field. However, this approach still faces two key challenges. \textbf{Challenge 1. Connectivity vs. Installation Cost:} LoRA performance is heavily affected by node transmission power and attenuation due to vegetation and terrain (\Cref{fig:figure2}), so achieving reliable wireless coverage demands detailed RF and power planning and often many gateways or towers to maintain the link budget across the target area, introducing a large upfront installation cost. \textbf{Challenge 2. Node Power vs Farm Operation Intrusive Form Factor:} Soil moisture sensing widely depends on active sensors (e.g. capacitive sensors consuming 10–30mW \cite{METER_TEROS12_2018}), and, when combined with long-range radios and the need for precise timekeeping to schedule wake cycles, these systems struggle to maintain an acceptable power budget. As a result, soil moisture sensing solutions can operate battery-free only with a large solar panel or otherwise require a large battery which is difficult to maintain in middle of the farm. Form factor tradeoffs between power needs and sensor performance lead to permanent fixture requirements, such as galvanized steel poles to support heavy solar panels and batteries \cite{vasisht2017farmbeats} or ground-based electrical conduits. These installations not only complicate deployment and increase cost but also obstruct farm vehicles during harvest and planting.

\begin{figure*}[!ht]
    \centering
    \includegraphics[width=\linewidth]{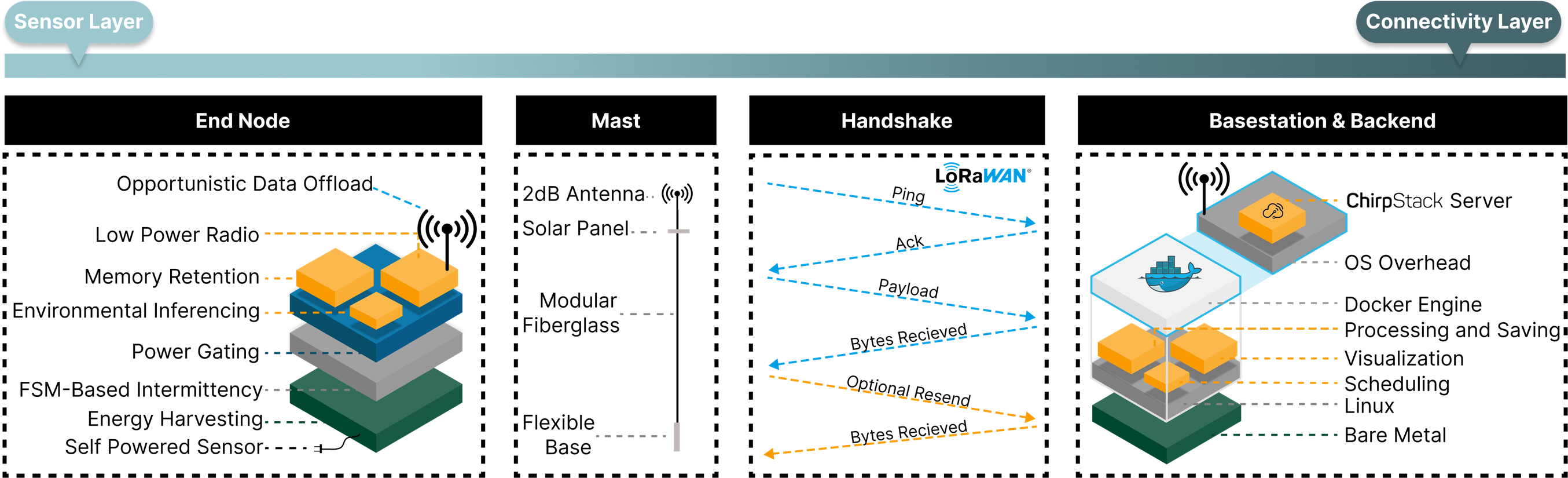}
    \caption{A block diagram of the sensor node , the activity layer, and their interactions.}
    \label{fig:figure3}
\vspace{-0.1in}
\end{figure*}

\noindent\textbf{Contributions.}  To address these challenges, we creatively introduce an end-to-end data collection platform that leverages the farmer’s existing daily movement as an integral part of the network and thus increases wireless coverage (\Cref{fig:firstfigure}). Rather than relying on permanent gateways or heavy-duty infrastructure, our system uses a mobile backend: a tractor, truck, or ruggedized golf cart (GATOR) equipped with a LoRa gateway that accompanies the farmer to the field edges during routine farm operations. This “data mule” opportunistically retrieves sensor logs whenever it comes within communication range of the sensor node, transforming intermittent connectivity from a limitation into a design feature. In a similar spirit, our node significantly reduces sensing power by leveraging typically written-off self-powered galvanic soil moisture sensor with an appropriate analog front end. Our system aligns the power and duty cycle of the sensor with the natural rhythms of farm work using a cohesive four-layer control framework (\Cref{fig:firstfigure}): a \textbf{Sensor Layer} controlled by a Finite State Machine (FSM), a \textbf{Connectivity Layer} that leverages FSM behavior to intelligently receive data, an \textbf{Activity Layer} that incorporates the farmer's activity and routes into the system, reminding them to visit areas near nodes whose data not been "picked up" in some days, and a final \textbf{Information Layer} that gives the farmer a data-rich display of field health. We evaluated the platform by measuring the node’s wireless coverage in a working corn farm in mid-season vegetation, testing the technical feasibility of self-powered soil moisture sensing across soil types representative of that field as well as characterizing  soil moisture sensing signal accuracy and power behavior under both day and night conditions. The platform supports the three contributions stated in the title:

\vspace{-0.15in}
\begin{enumerate}[leftmargin=10pt] 
\item \textbf{Low-power.} The battery-free node operates at only 180~\textmu W, providing wireless soil moisture and temperature sensing, reducing energy harvester to small form factor solar cell. It achieves this through: (a) \emph{Self-powered sensing.} Our system uses a high-impedance buffer, self-powered galvanic cell as the soil moisture sensor, eliminating the need for power-hungry capacitance/high-frequency-based sensing while co-locating temperature measurement. (b) \emph{Adaptive runtime behavior.} Powered solely by solar harvesting into a single storage capacitor, the system conserves energy during cloudy or dark periods and prioritizes transmission during sunny periods. With an optimized wake-up strategy, the device can fully recharge in one cycle and operate through extended periods of darkness ranging from 21 days using a 20-minute timer, down to 5.2 days on a 5-minute timer (Table 2).

\item \textbf{Activity-aware.} Our platform is designed not to interrupt farm activity, but to leverage it.
(a) \emph{Connectivity from farmer mobility.} Our system inverts the traditional networking model by treating the farmer’s daily routes as the source of mobility. Routine passes with tractors, trucks, or GATORs form the network fabric, keeping RF transmission power overhead low, extending node power lifetime, and avoiding the cost of permanent infrastructure.
(b) \emph{Flexible, modular mast.} A multi-section fiberglass mast co-locates the radio antenna and solar panel and supports communication ranges up to 1 km in growing agricultural fields without a permanent fixture (\Cref{fig:firstfigure}). The mast is mounted on a metal spring, height-tuned to the target crop and typical tractor clearance, and placed in-line with the crop so that it bends out of the way just like any other crop in the field when equipment passes over it.

\item \textbf{Low-cost.} A single node costs less than \$35 (Table 3) and the mobile base station costs less than \$100. With one base station and 20 nodes, a 100-acre deployment requires an upfront investment of under \$675, with no additional infrastructure, recurring costs, or maintenance overhead.
\end{enumerate}
\vspace{-0.05in}
The platform fits within existing farm operations and budgets offering a path to economic gains through improved yield and input efficiency without imposing new recurring costs on the farmer, and supporting more sustainable and economically resilient agriculture in both high- and low-income settings \cite{sanyaolu_role_2024}.

\begin{figure}[b]
\vspace{-0.25in}
    \centering
    \includegraphics[width= 0.95\linewidth]{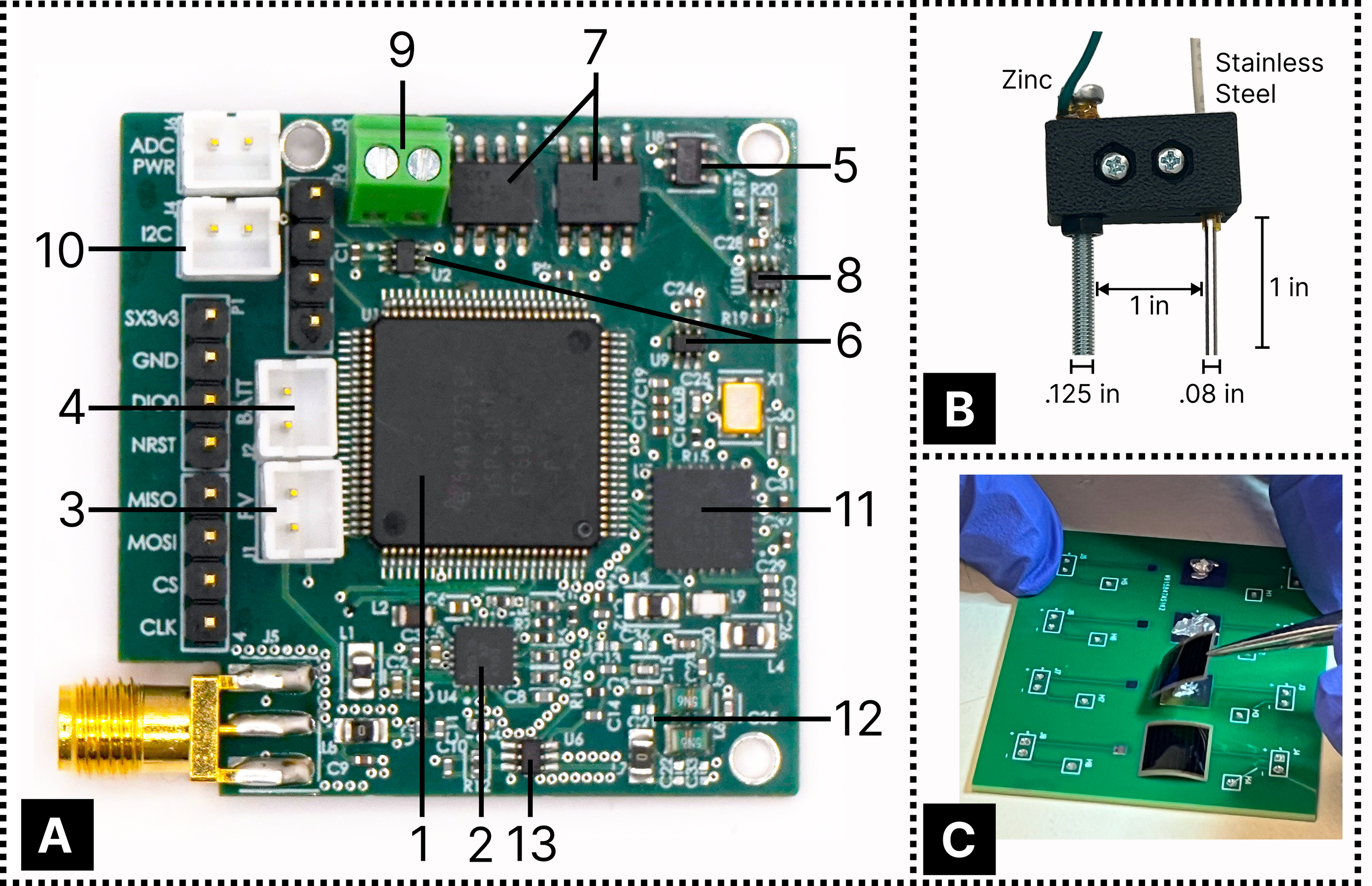}
    \caption{(A) The wireless sensing node PCB representing the end node as in the block diagram of \Cref{fig:figure3}, with labels described in  \Cref{sec:sensor_layer}. (B) Self-powered galvanic soil moisture sensor (\Cref{sec:sensor}). (C) Manufacturing of the $2x1{cm^2}$ solar panel connected in series on the development board.}
    \label{fig:figure4}
\end{figure}

\section{Platform Design}
\noindent Our system is a soil moisture sensing platform that balances cost, power, connectivity (range and base station availability), and a farm-operation-friendly form factor. It is defined by four distinct layers, each designed with the behavior of the others in mind to enable seamless integration into normal farm operations and to provide intuitive soil moisture data to the farm operator. We describe the technical details and operation of each layer below. 
\vspace{-0.1in}
\subsection{Sensor Layer}\label{sec:sensor_layer}
The sensor layer consists of  (\Cref{fig:figure3} left) three main parts: 
(i) a buried soil moisture sensor  (\Cref{fig:figure4}B, \Cref{sec:sensor}) (ii) an end node PCB consisting of an energy harvester, LoRA radio, and timekeeping circuit (\Cref{fig:figure4}A,  \Cref{sec:energy_harvester}-\Cref{sec:radio}) (iii) and a minimal form factor antenna mast (\Cref{fig:figure3}, \Cref{fig:figure12}, \Cref{sec:mast})
Below, we explain the key design decisions for selection of each component.  

\subsubsection{\textbf{Self-powered Galvanic Soil Moisture Sensor}}\label{sec:sensor}\hfill
\noindent\textbf{Device design.} The soil moisture sensor consists of two metal wires, one made of zinc (.125-inch diameter) and another of stainless steel (.08-inch diameter). Each has a length of 1 inch and they are placed 1 inch apart in a 3D-printed frame (\Cref{fig:figure4}B).

\begin{figure}[H]
\hspace{-10pt}
    \centering
    \includegraphics[width=0.9\linewidth]{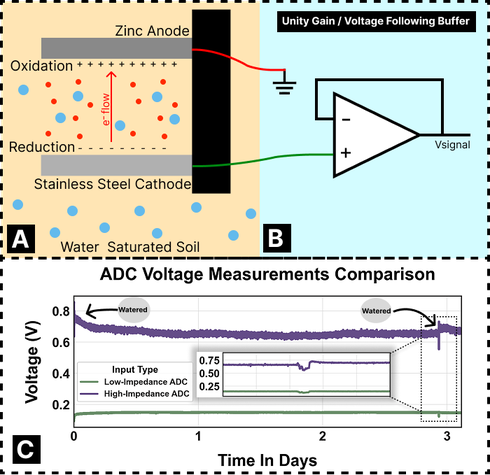}
    \caption{(A) Galvanic cells naturally form when two dissimilar metals form a closed circuit, with the cathode losing electrons to the anode through some medium (in our case, water-saturated soil). (B) Our system uses high-impedance circuitry to buffer the cell, preventing multimeter loading. (C) Preliminary experiment comparing galvanic cells with high- and low-impedance front ends. Note the low impedance sensor's poor response during a watering cycle.}
    \label{fig:figure5}
\end{figure}

\noindent\textbf{Theory of Operation.} A galvanic cell is formed when two dissimilar metals with different electrode potentials—such as zinc and stainless steel—are electrically connected while sharing an ionic medium like water-permeated soil (\Cref{fig:figure5}A). The metal with the more negative reduction potential (zinc) oxidizes, releasing electrons and forming $Zn^{2+}$ ions, making it the anode; stainless steel, being more noble, acts as the cathode and accepts electrons. Electrons flow through the metal connection from zinc to stainless steel, while ions in the soil move to maintain charge balance—$Zn^{2+}$ diffuses into the soil, and dissolved oxygen typically drives the cathodic reduction reaction (often forming $OH^{-}$). The galvanic cell can also be viewed as a self-powered soil moisture sensor. As the water ions in the soil increase, so does the voltage output.

\noindent\textbf{Gap in Adoption of Galvanic Sensor.} Galvanic-cell-based soil moisture sensors have a reputation for lacking durability and being susceptible to drift over time as they quickly corrode. This reputation is largely due to an "always-on" state of the cell itself, with the reaction between the anode and the cathode driving irreversibly forward until the reactants are consumed.  

\noindent\textbf{Our Solution: High-Impedance Buffer.} To leverage the self-powered nature of the galvanic soil moisture sensor while maintaining usable operation, we use a high-impedance buffer as the analog front end (\Cref{fig:figure5}B). Buffering the cell with a high-impedance input minimizes multimeter loading, where the measurement instrument itself acts as a load on the circuit. By using a high-enough  impedance ($>$10x the internal resistance), the measurement circuitry places effectively no load on the cell, measuring strictly the potential of the cell and minimizing the forward reaction rate. This method produces clearer and higher voltage measurements (\Cref{fig:figure5}C) and slows sensor deterioration, extending sensor life and presenting a viable method of passive sensing for low-power devices (\Cref{sec:longevity}).

\vspace{-0.1in}

\vspace{-0.1in}
\begin{figure}[!ht]
    \centering
    \includegraphics[width=\linewidth]{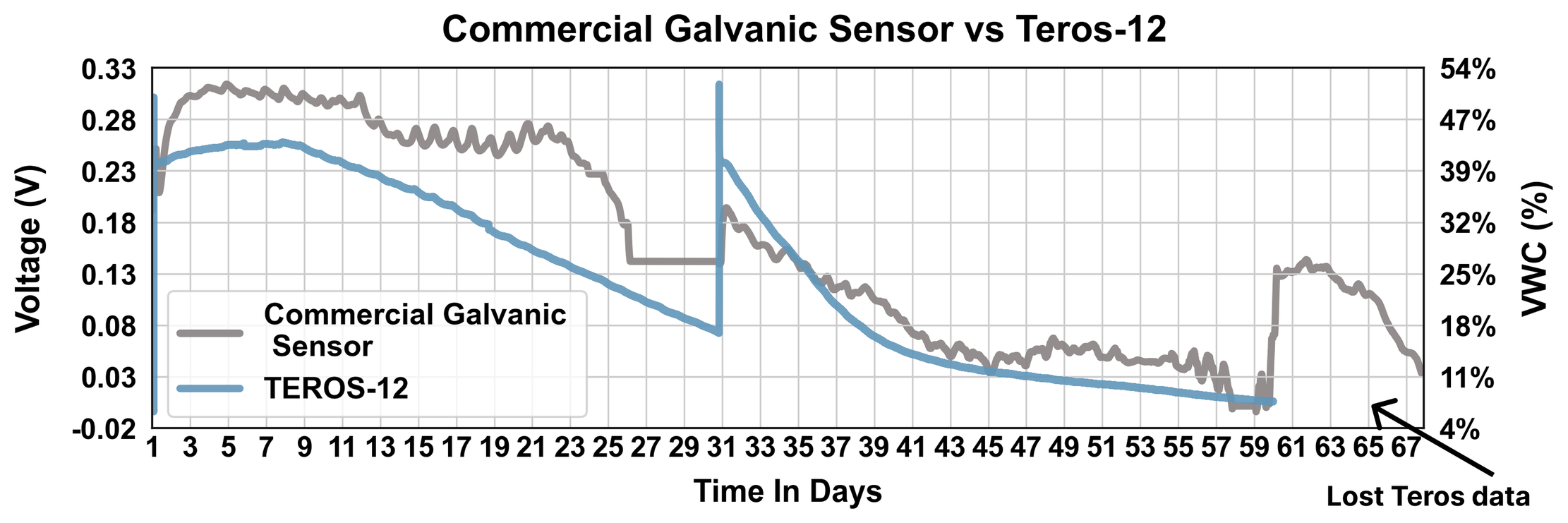}
    \vspace{-0.3in}
    \caption{Preliminary experiment of commercial zinc/aluminum galvanic sensor probe against TEROS-12 \cite{METER_TEROS12_2018}.}
    \label{fig:figure6}
     
\end{figure}

\noindent\textbf{Preliminary Experiment.}
We conducted preliminary lab experiments by using the anode and cathode of a commercially available zinc/aluminum sensor (probe from \cite{irtov_soil_meter_amazon} without front end) in combination with a 10 GOhm impedance data acquisition logger from National Instruments \cite{ni_entry_level_usb_daq} (\Cref{fig:figure6}). The galvanic sensor demonstrated signal similarity to a commercial capacitive sensor \cite{METER_TEROS12_2018} over 70 consecutive days of readings across two wet–dry cycles. Inspired by this result, we built our own PCB and custom galvanic sensor.

\noindent \textbf{Analog Sensor Front End.}
In order to achieve high impedance and prevent the galvanic cell-as-sensor (\Cref{fig:figure4}, label 7 and B) from consuming itself, the node employs an OPA-192 op-amp (\Cref{fig:figure4}, label 9) in unity gain voltage follower mode. The characteristic low quiescent current of this op-amp results in an input impedance of 1 TOhm to buffer the sensor.

\noindent\textbf{Sensor Material.}
Based on galvanic cell theory \cite{feiner1994nernst}, the electrode materials directly affect the voltage produced. In our preliminary experiments, a zinc/aluminum pair generated only about 200 mV at 32\% volumetric water content (VWC), so we instead explored a zinc/stainless steel pair for our custom sensor. This combination offers better corrosion resistance, low cost, and, due to the relative positions of zinc and stainless steel in the galvanic series, produces a larger signal of approximately 600-700 mV (\Cref{fig:figure6}).

\noindent\textbf{Sensor Spacing.}
We further deploy our sensor and evaluate different types of soil collected from a real corn field in \Cref{sec:evaluation}. Since the distance between the electrodes affects the voltage magnitude and the resistance of the soil medium between them varies approximately linearly with spacing, we fix the electrode separation at 1 inch. This controlled spacing allows direct comparison across samples while still providing a reasonable volume of soil and moisture between the electrodes. 

\subsubsection{\textbf{Energy Harvesting}}\label{sec:energy_harvester}\hfill

 \noindent \textbf{Solar Panels.}  (\Cref{fig:figure4}, label 3) For our node, we chose two MicroLink Devices $1cm^2$ triple-junction inverted metamorphic (IMM) solar cells connected in series for their superior energy density performance \cite{microlink_photovoltaics}. The total area is only $2cm^2$, which does not disrupt the form factor or add weight. 
 
\noindent \textbf{Texas Instruments  BQ25570 Power Management.} (\Cref{fig:figure4}, label 2) The Texas Instruments BQ25570 power harvesting unit was chosen for its high reliability in charging batteries and capacitors from a solar panel. The device pulls 8-14 mW from our solar panels in direct sunlight and ensures the capacitor has sufficiently charged before enabling the MCU to prevent brownout events from inrush current. Equipped with Maximum Power Point Tracking (MPPT) algorithms, the device functions in easily identifiable patterns under different lighting conditions, allowing the FSM to make quick power availability decisions. 
 
\noindent \textbf{5.5V 1 F Supercapacitor.} (\Cref{fig:figure4}, label 4) The node uses a Kemet low-leakage supercapacitor with an energy overhead given:
\begin{equation}
    E = \frac{1}{2}C(V_{max}^2-V_{min}^2)=\frac{1}{2}(1F)[(5.5V)^2-(3.3V)^2]=9.68J.
\end{equation}

A detailed evaluation of energy harvested under varying light intensities,including charge life, is presented in \Cref{sec:evaluation}.


\subsubsection{\textbf{Power Gating and FSM-Based Intermittency}}\hfill \\
The node is governed by a duty-cycled FSM to produce predictable behavior and maximize inter-layer compatibility. The components used are described below. 

\noindent\textbf{Microcontroller.} An MSP430FR6989 (\Cref{fig:figure4}, label 1) is employed for default-off intermittent node operation for the following reasons: (i) During duty cycling, the ultra-low shutdown current of 20 nA, together with 128 kB of on-board FRAM, outweighs the drawback of the MSP430's legacy hardware drawing 0.8 mA in active mode. (ii) The on-board 12-bit ADC provides sub-mV resolution for our voltage measurements, which oscillates only between 0 and 1V. (iii) We gain multiple SPI and I2C interfaces for additional temperature sensors and peripherals. (iv) The MCU natively supports AES-128 encryption, mandatory for LoRaWAN network connectivity.

\noindent \textbf{Low-Power External Timer.} Texas Instruments TPL5111  (\Cref{fig:figure4}, label 5) governs the node's intermittent wake-ups for exiting MSP430's Low-Power Mode (LPM) 4.5. Consuming only 35 nA, the device triggers interrupts for our MCU with a resistor-divider-programmed period ranging from 100\,ms to 2\,hours. This enables the minuscule 20\,nA current draw in LPM4.5, an improvement of 0.1 \,mW alone from departing from LPM3.5. LMP4.5, however, allows us to forego usage of a 32 kHz external oscillator normally present with the MCU, saving an additional 16 mW. 

\noindent \textbf{Low-Power Load Switch.}
We use a TPS22919DCKT  (\Cref{fig:figure4}, label 6), which consumes just 8 uA in active operation and 2 nA when off, in order to minimize leakage current into inactive components. The node power-gates the SX1276 radio module while not transmitting. We similarly gate power to the OPA-192 op-amp, our most power-hungry device at 1\,mA. 

\vspace{-0.05in}
\subsubsection{\textbf{Low-Power Environmental Inference}}\hfill

\noindent \textbf{Light Level.}  To infer environmental conditions (e.g., sunny/cloudy, day/night), the sensor layer employs current and voltage sensing with the solar panel. For voltage, the node employs a power-gated OPA-192 (\Cref{fig:figure4}, label 7) to the internal ADC.  For current sensing, the node uses an INA212AIDCKR  (\Cref{fig:figure4}, label 8). Detailed characterization is done in \Cref{sec:evaluation}. 

\noindent \textbf{Temperature Sensor.} We employ a miniature, ultra-low-power ST-40 from Sensirion (\Cref{fig:figure4}, label 10). Using I2C for ease of communication, this IC sits off the board and in the soil with the electrode, consuming just 0.4 uA in active mode, and is similarly power-gated when not in use.

\vspace{-0.05in}
\subsubsection{\textbf{Non-Volatile Memory Retention}} \hfill

\noindent The MSP430 natively supports extremely low-power writes to its FRAM, enabling seamless integration into the FSM. 

\vspace{-0.05in}
\subsubsection{\textbf{Low Power Radio}}\label{sec:radio}\hfill

\noindent \textbf{LoRA radio and RF Front End.} The node uses a Semtech SX1276 transceiver (\Cref{fig:figure4}, label 11) for LoRa-based wireless communication. The SX1276 is the underlying chip to many commercially available chiplets, including the commonly used REYAX RYLR998, with the added benefit of communicating directly with the chip via SPI as opposed to requiring a middle layer translating UART commands. Standard LoRa system-on-modules available for purchase generally use the highly flexible 137 MHz to 1020 MHz output channel as opposed to the narrower, more power-efficient dedicated 915 MHz channel \cite{Semtech2016_SX1276}. Custom PCB design afforded the ability to use this more efficient channel, as well as disconnect the power amplification pin on the radio.
The chip is configured for the standard US 915 MHz and affords enough flexibility for either plain-text or encrypted transmission, which enables compatibility with commercially available LoRaWAN services such as ChirpStack and The Things Network. 

\noindent\textbf{pSemi 4259 RF Switch.}  The pSemi 4259 RF switch (\Cref{fig:figure4}, label 13) was chosen for its usage in SX1276 evaluation kits. Employing these components on a custom PCB required a custom-tuned impedance matching circuit  (\Cref{fig:figure4}, label 12) in order to keep the bare SX1276 chip instead of a commercially available product, which would have cost valuable active-time power consumption.

\noindent \textbf{Antenna}: We use a 2 dBi 915 MHz omnidirectional half-wave dipole whip antenna, attached to the PCB through an SMA connector and coaxial cable for weatherproof installation.

\vspace{-0.05in}
\subsubsection{\textbf{Flexible Fiberglass Mounting Point}} \label{sec:mast} A narrow, lightweight, 4-meter fiberglass rod is leveraged to deploy the node (\Cref{fig:firstfigure}, \Cref{fig:figure3}, and  \Cref{fig:figure12}). The rod elevates the antenna and low-profile solar panel mount above the Fresnel zone and into direct sunlight. The rod is mounted to a flexible metal spring, maximizing the mast's flexibility so it can clear farming equipment such as tractors or even be run over and spring back upright once the vehicle has passed (\Cref{fig:figure12}).
The rod comes in 4 separate pieces, allowing operators to easily carry a node into the field in their vehicles and tailor the height to their crop of choice. This flexible mast, combined with a small energy harvester enabled by the platform's low power budget, allows each node to maintain a form factor that does not interfere with routine farm operations. 
\vspace{-0.1in}
\subsection{Connectivity Layer.}
Our platform inverts the traditional networking model: rather than extending network infrastructure, it treats the farmer’s existing movement patterns as the substrate for reliable data delivery. A vehicle-mounted LoRaWAN gateway passively collects data while accompanying routine fieldwork. This mobility-aware backend design allows sensor nodes to transmit infrequently and at low power, relying on scheduled, predictable proximity to the gateway for reliable uplink confirmation.

\noindent\textbf{Low Power Connectivity} LoRa transmissions are generally afforded a capability of several kilometers, and even up to 15 km \cite{lora_range}. A lack of reporting on power consumption during these transmissions required experimentation to fully understand the range/power tradeoffs. Additionally, we sought to understand the severity of attenuation by crops with antennas placed at different heights above or within the field. Of particular concern was the \textbf{Fresnel zone}, the ellipsoidal region around the direct line of sight between two antennas in which multipath interference is non-destructive. The radius of this region is given by
\vspace{-0.15in}

\begin{equation}
    R = \sqrt{\frac{\lambda \, d_1 \, d_2}{d_1 + d_2}}
\end{equation}

where $\lambda$ is the wavelength, and $d_1$ and $d_2$ are the distances from an 
obstruction to the two antennas. This region is the most conducive to robust transmission, so it must remain as free of obstructions as possible. Exploratory deployments for optimization of this concepts are completed in Section 3.3.

\noindent\textbf{Native Host Processing and Dockerized ChirpStack Gateway}
The gateway is designed as an easy-to-implement service capable of running entirely locally. The backend containerizes the LPWAN gateway networking services; by running the ChirpStack network and application servers within Docker containers, the system can be deployed on any low-power single-board computer or existing farm PC without dependency conflicts. This containerization also enables the LPWAN backend, normally reliant on an internet connection, to be operated entirely locally, providing full functionality from the GATOR itself. Meanwhile, all data filtering, post-processing, and visualization services run natively on the host operating system to ensure simple integration with ChirpStack services. Such modularity allows the farmer to own the entire data pathway and avoid reliance on external connectivity.

\noindent \textbf{Opportunistic Data Offload}
By minimizing transmission power and heavily duty-cycling operation, a final design consideration must be met: coordinating a time and place for transmission to occur successfully. Our system combines a handshake-based exchange strategy with deterministic transmission behavior and a well-defined propagation range to create predetermined pickup zones for a data mule to receive data. These pickup zones are chosen for their overlap with other nearby nodes and their proximity to normal farming operations, so that a data mule co-located with farming equipment performing regular activities might be able to passively collect data and therefore minimize disruption of day-to-day operations.

\noindent \textbf{Coordinated Intermittent Communication} Sensor nodes operate under a fixed 20-minute wake cycle governed by a low-power external timer. Upon waking, the node assesses its energy state and determines whether to initiate transmission. If energy is sufficient, it initiates the handshake with the base station and awaits a response from a gateway within range. This minimal protocol ensures that data is only marked as “delivered” if explicitly acknowledged. The node retains the reading in FRAM for later retry, allowing for 4600 cycles at 9 bytes per cycle (63 days of capacity at 20-minute cycles). Because timing behavior is deterministic, both successful and missed exchanges are predictable, enabling reliable delivery across multiple gateway passes without maintaining a persistent link or incurring idle listening cost.

\noindent\textbf{Gateway Hardware and Networking} Our mobile gateway consists of a lightweight Linux workstation running a containerized ChirpStack LoRaWAN server and a WM1302 concentrator module. It is powered by a rechargeable USB battery and uses the same flexible 4-meter mast antenna as the sensor nodes. This setup provides wide-area reception while remaining compact and easy to mount on existing equipment. The gateway operates entirely offline—packet decryption, acknowledgment generation, and queue management are handled locally, eliminating reliance on cloud infrastructure or persistent internet connectivity. With the ability to support thousands of simultaneous LoRa messages, this mobile platform scales effectively with field size and sensor density.

\vspace{-0.15in}
\subsection{Sensor \& Communication Layer Interaction}
\textbf{Embedded Software.} The core functionality of this data acquisition system is to use context-aware methods to conservatively enter predefined power modes, each optimized to expend the least amount of energy necessary to collect or transmit data to an edge-connected base station. Placing the highest priority on data preservation, we minimize on-board data processing, and govern our power-aware context through the employment of an FSM. 

\textbf{FSM-Based Intermittency}
By optimizing for extreme power budgets and embracing intermittency as a design feature, the device needs to behave deterministically when communicating with the backend in order to conserve power, ensure predictability, and have the ability to recreate timestamps for data analysis. By leveraging the hardware-defined wake-up schedule and environmental inferencing, the system can also make smart decisions on how to best preserve power. Since neither sunlight nor nearby farming equipment are available during nighttime hours, we preserve vital power during periods we know radio transmissions will not be successful by operating in a simple FSM. During each boot cycle, the node determines how much power is available based on context inferred through the solar panel sensing.

\begin{figure}[H]
    \centering
    \includegraphics[width=\linewidth]{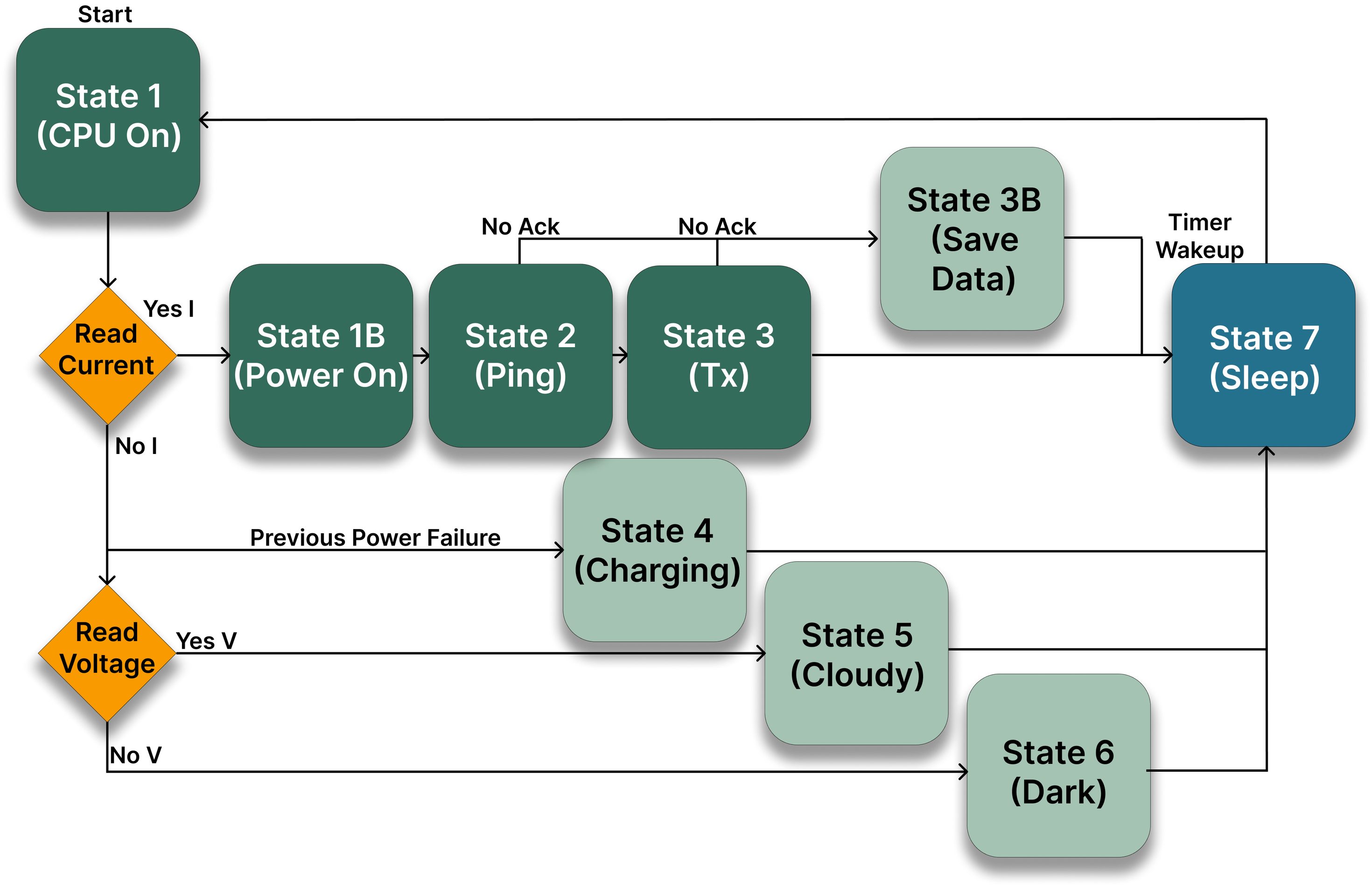}
    \vspace{-0.2in}
    \caption{A general flow diagram of the FSM.}
    \label{fig:figure7}
\end{figure}

\noindent\textbf{FSM} The system governs sensing, operation, and transmission in order to optimize energy consumption. To maximize energy efficiency, the FSM begins in a baseline state that only verifies power thresholds before branching into subsequent operational phases (\Cref{fig:figure7}). Each state is designed to activate only the minimal set of peripherals necessary for its task—such as ADC sampling, FRAM writes, or radio transmission—before promptly transitioning onward. In this way, the FSM partitions behavior into states corresponding to sunlight conditions (sunny, cloudy, or dark), data handling (buffering, transmission, and acknowledgment), and deep sleep (LPM4.5). This structure ensures predictable control flow while aggressively minimizing power consumption. 

\vspace{-0.20in}

\begin{figure*}[t]
    \centering
    \includegraphics[width=\linewidth]{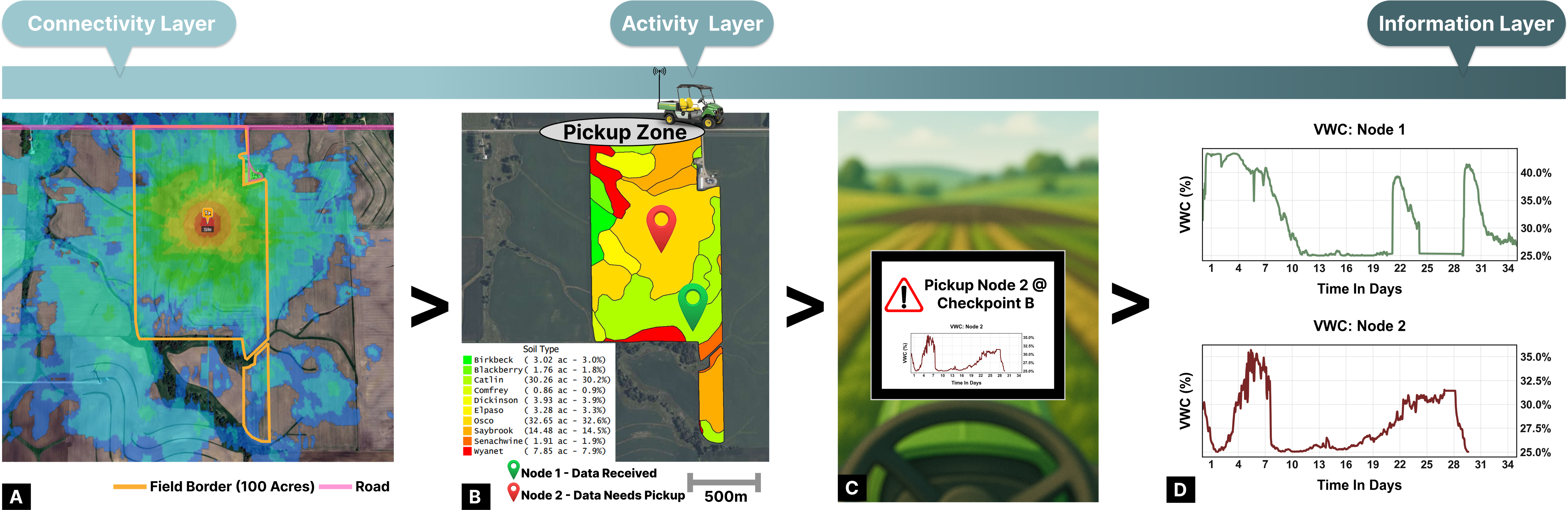}
    \caption{Proposed tool workflow. (A) shows propagation overlays of the 915MHz spectrum at our system's specifications. The farmer can use this info overlayed with existing roadways, coupled with knowledge of soil maps (B) to predefine pickup zones This view also indicates contact recency from the nodes.  With the basestation running entirely locally, a portable dashboard can recommend schedule changes to the farmer (C), or allow them to view granular node data (D).}
    \label{fig:figure8}
    \vspace{-0.1in}
\end{figure*}

\noindent\textbf{State 1 (CPU On):} The entry point of the FSM. The system checks for available current and voltage and infers charging conditions before determining the appropriate branch.

\noindent\textbf{State 1B (Power On):} The system annotates the result as \emph{sunny}, powers on the radio, and enters State~2. 

\noindent\textbf{State 2 (Ping Base Station):} The system pings the base station. If an acknowledgment is received, the FSM enters State~3. Otherwise, the FSM enters State~3B.

\noindent\textbf{State 3 (Transmit Data):} The current data, as well as any data in the FRAM buffer, is transmitted. If an acknowledgment is received, the FSM enters State~7. Otherwise, the FSM enters State~3B.

\noindent\textbf{State 3B (Save Data):} The system writes data to the FRAM buffer and proceeds to State~7.

\noindent\textbf{State 4 (Charging):} State 4 occurs only if the radio knows it is coming out of a period of darkness where the solar panels have not been charging the capacitor. If the state 6 "dark" flag is set, it powers on the radio and informs the base station with a specific ping notifying that it is still alive but data comingon \textit{the next} cycle to allow for charging. It samples the ADC, then proceeds to State~7.

\noindent\textbf{State 5 (Cloudy):} The system recognizes a period of diffuse light not optimal for charging. It samples the ADC, annotates the result as \emph{cloudy}, writes data to the FRAM buffer, and proceeds to State~7.

\noindent\textbf{State 6 (Dark):} The system recognizes a period of darkness and is not charging. It samples the ADC, annotates the result as \emph{dark}, writes the data to FRAM, and transitions to State~7.

\noindent\textbf{State 7 (Low-Power Sleep):} The system enters deep sleep mode (LPM4.5) until woken by an external timer interrupt, after which the FSM resumes at State~1.

The most important state of the FSM is State~1, in which power availability is sensed and which serves as the default state for the machine. This state is triggered only once every 20 minutes by a GPIO interrupt from an external timer. The MCU boots up and turns on just one ADC peripheral, successively reading from 2 points in our PCB in turn to understand if the solar panel is charging the device actively and thereby inferring the state of the capacitor charge. This gives the device the ability to configure its tasks for a given runtime as well as annotate the environmental conditions (namely whether it is sunny, cloudy, or dark). Regardless of state, every state logs several data points: sun state, ambient temperature, and signal voltage. All signal processing to correct for noise or temperature is done off-device. 
\begin{figure*}[t]
    \centering
    \includegraphics[width=\linewidth]{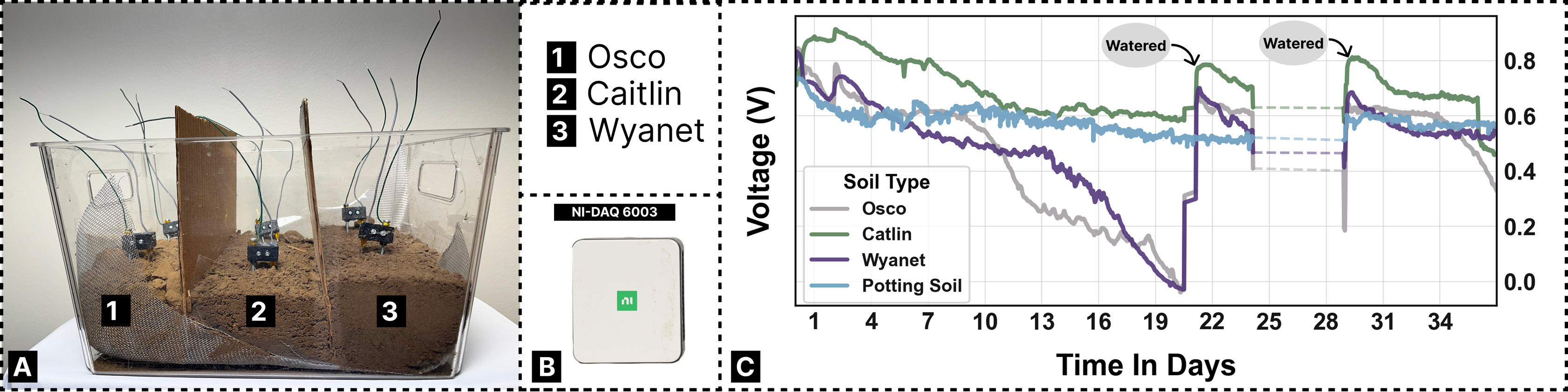}
    \caption{(A) Soil samples from a local farm. (B) NI-DAQ 6003 high-impedance data logger. (C) The resulting voltage data from the samples. The samples were logged with equal amounts of volumetric water content to begin and subsequently watered two more times throughout the test. Note the varying rates of drying in the soils, which ranged from well draining silt to poorly draining clay.} 
    \label{fig:figure9}
\vspace{-0.15in}
\end{figure*}

\subsection{Activity Layer.}  
The Activity Layer must abstract the physical field into an easy-to-interpret dashboard that allows the farmer to seamlessly interact with the wireless sensor network. It needs to interpret inputs from the previous layers to determine where nodes are located, how long it has been since data was last collected, whether intervention is necessary, and where a vehicle should pause or traverse to retrieve information efficiently.

\noindent\textbf{Farmer Activity as Connectivity} Our system is designed to operate symbiotically with the farmer’s routine. The gateway is affixed to vehicles already in use—such as tractors, trucks, or carts—without requiring changes to travel paths, schedules, or tasks. This enables seamless integration into the farming workflow: data is collected as a side-effect of normal motion through the field. The farmer is not responsible for operating the gateway, interpreting status messages, or initiating communication. Instead, the system minimizes cognitive and behavioral overhead, embedding sensing into natural mobility rather than imposing infrastructure onto the field.

\noindent\textbf{Visualization and Scheduling} Passive data collection by data mules from batteryless nodes will require daily confirmation that data collection is indeed occurring. To ensure robust data collection from intermittently reachable sensors, we provide a visual tool that tracks the last successful contact with each node. A simulated field can be viewed in \Cref{fig:figure8}A, \Cref{fig:figure8}B. Nodes are color-coded by recency: green (within 24 hours), yellow (1–3 days), and red (more than 3 days). The interface overlays these status markers on a field map, highlighting our predetermined pickup zones--locations where brief pauses during regular activities enable reliable synchronization. This feedback allows the farmer to recover missed transmissions opportunistically, without needing to manage or directly interact with the network. In addition to contact recency, we add the ability to view raw data from each node, enabling an at-a-glance dashboard of what is happening in the field, and whether or not operator intervention is required (\Cref{fig:figure8}C, \Cref{fig:figure8}D). We implemented this visualization functionality in Python for our local corn field. Overlays are built manually, leaving automation and path planning to future work. This tool thus bridges the Connectivity and Information layers. 

\vspace{-0.1in}
\subsection{Information Layer.}  
The information layer must capture and reuse the data already present in the farm’s digital ecosystem. This includes map overlays of the field, soil-type metadata, yield histories, and other information routinely gathered through existing operations. Integrating these data sources into the tools ensures that sensor placement and calibration are informed by the farm’s own historical record rather than external assumptions. For example, the soil map appearing in \Cref{fig:figure8}B is courtesy of our local farm management, and shows how existing tools can be overlayed into our visualization to help view, plan, and understand field dynamics.

\textbf{Local Data Logging and Visualization} All received packets are decoded and authenticated using ChirpStack’s standard LoRaWAN protocol stack at the base station. A lightweight MQTT listener records each successful uplink to a time-stamped CSV log. This data stream is visualized directly on the dashboard used by the farmer, providing real-time insight into sensing coverage and battery-less node health. The system requires no server infrastructure, cellular connectivity, or wireless backhaul. All processing occurs locally on the mobile gateway, enabling use in remote, disconnected environments while preserving compatibility with widely adopted LoRaWAN tooling.

\section{Evaluation} \label{sec:evaluation}
We evaluate the system across two primary dimensions: galvanic sensor performance and power consumption.

\vspace{-0.1in}
\subsection{Soil Moisture Sensor Characterization}
\subsubsection{\textbf{Sensor Feasibility}}\hfill

\noindent We evaluated the sensor feasibility by signal mapping of our  galvanic soil moisture sensor voltage against a TEROS-12 probe—an industry-standard, high-precision instrument as ground truth\cite{METER_TEROS12_2018}. 

\noindent\textbf{Understanding TEROS Measurements.}
The TEROS-12 measures the dielectric permittivity of the soil by supplying a 70-MHz oscillating wave to its sensor needles. The sensor microprocessor determines the charge time, which is proportional to the substrate's dielectric permittivity, and outputs a raw digital value representative of this property.
TEROS provides RAW values that need to be mapped to VWC. We employ the Topp equation \cite{topp_electromagnetic_1980}, an estimate of soil water content without requiring soil-specific calibration, linking the dielectric permittivity of a soil to its Volumetric Water Content (VWC: the ratio of water volume to soil volume). This equation has become the standard reference for the calibration used by most commercial sensors, including the TEROS-12 manual, which provides the following calibration equation to convert its sensor output directly to VWC \cite{METER_TEROS12_2018}: 
\vspace{-0.1in}
\begin{equation}
 \Theta(\frac{m^ 3}{m^3} ) = (3.879E^{-4} * RAW-0.6956) 
\end{equation}
where $\Theta$ represents the VWC and RAW represents the TEROS-12 sensor output. For interpretability, VWC can be expressed as a percentage: 
\vspace{-0.15in}
\begin{equation}
  \Theta(\%)=\Theta(\frac{m^ 3}{m^3} ) * 100 
\end{equation}

\noindent\textbf{Experimental Setup.} Due to the TEROS-12’s active nature, cross-calibration against our passive probe is non-trivial. Any passive sensors placed within the same pot as the TEROS-12 yielded non-usable results, and need to be electrically isolated. Thus, for this test, two identical pots were placed in the same indoor lab environment, with identical soil types and amounts of water added. One pot housed three of our galvanic sensors (Zn/SS) in addition to one commercial sensor (Zn/Al) and the other pot housed the TEROS-12 as ground truth. To ensure measurement accuracy, our galvanic sensor was measured with the 10 GOhm impedance NI-DAQ 6003 (\Cref{fig:figure9}B) at a sampling rate of 1 Hz. 

\noindent\textbf{Signal Mapping Steps.}
In order to obtain VWC from our sensor's (Zn/SS) output, we needed to establish a mapping function from our sensor's raw voltage values to TEROS-12 raw sensor outputs and then convert to VWC percentages using Equations 3 and 4. 

\begin{figure}[!ht]
\vspace{-0.15in}
    \centering
\includegraphics[width=\linewidth]{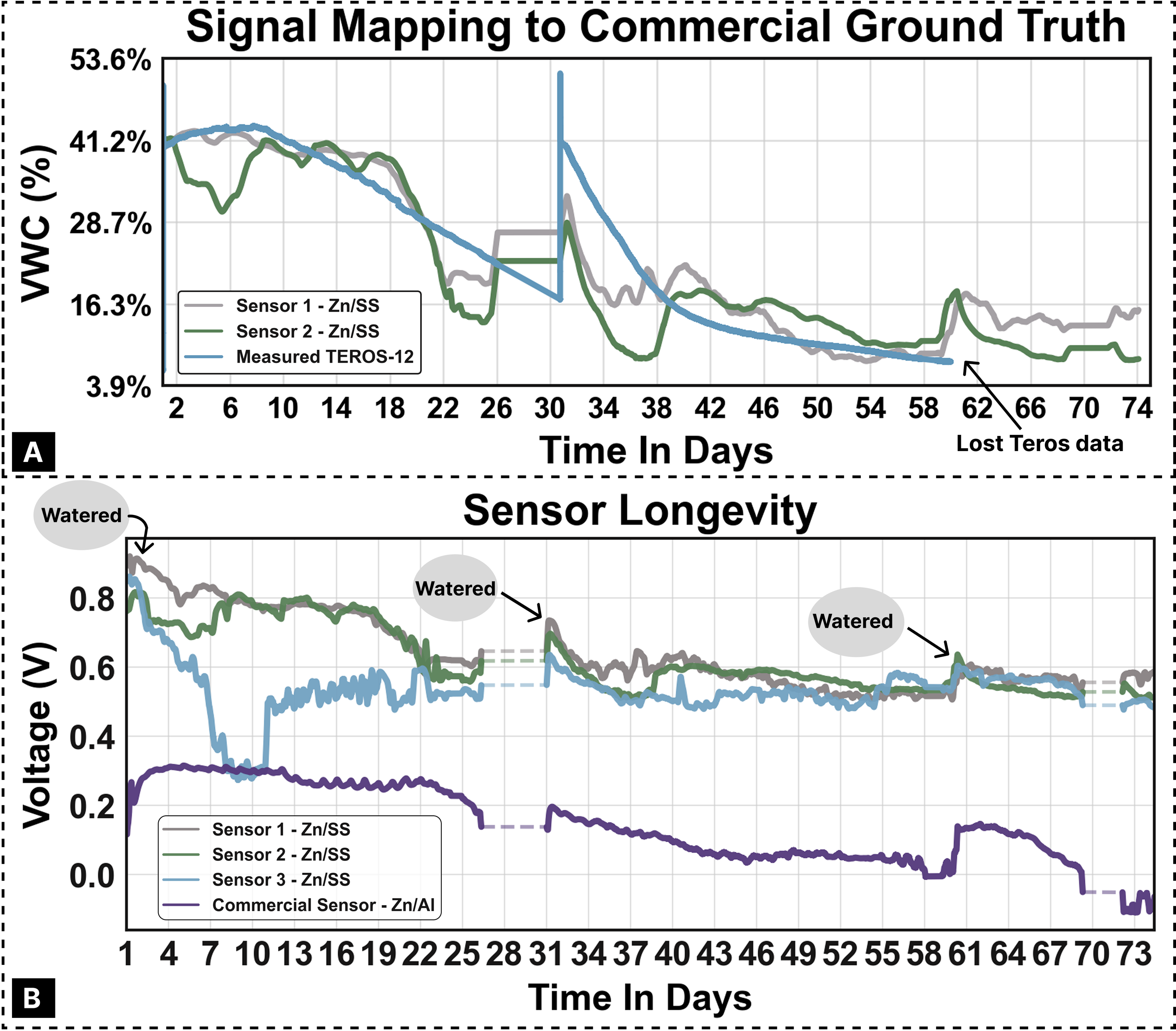}
    \caption{(A) Our sensor's VWC values (raw voltage mapped to TEROS-12 raw values and then converted to VWC) against the TEROS-12's VWC (raw output converted to VWC). (B) Our sensor's data collected over 70 days.}
    \label{fig:figure10}
\vspace{-0.15in}
\end{figure}

We used 73 days of data from sensor 1 depicted in \Cref{fig:figure10}B to create a training subset. This training subset was first used to test several regression models, using our sensor 1's voltage and the RAW output from the co-located TEROS-12 as parameters. The cubic polynomial model performed best, providing a coefficient of determination ($R^2$) of \(85.8\% \) and a prediction error of $\pm 9\%$ across the range of ground truth VWC values. It outperformed both the linear and quadratic models which provided an $R^2$ of  \(83.0\%  \) and \(84.9\%\) respectively.  Model robustness was further validated using a 10-fold cross validation with a 90/10 train-test split. Across all folds, the cubic model consistently yielded an average prediction deviation of $9 \pm 0.01\%$. The resulting calibration equation was: 
\begin{equation}
 RAW = -2.34 * 10^4 V^3 + 4.45 * 10^4 V^2 - 2.46 * 10^4 V + 6.09 * 10^3
\end{equation}

Applying this equation to sensor 2's voltage outputs yielded an $R^2$ of \(70.64\% \) with an average prediction deviation of  $\pm 13.9\%$. Sensor 1 and sensor 2's resulting VWC percentages compared to the TEROS-12's VWC percentages are depicted in \Cref{fig:figure10}A.

Both visually and numerically, this performance suggests variations due to deviations in soil content within the pot, non-uniform watering, and susceptibility to ionic activity within the soil itself (i.e. nutrient location). In future work, we will develop a more rigorous data collection and mapping algorithm, using measurements from sensors across diverse soil types, to more accurately relate sensor voltage to volumetric water content (VWC). 

\noindent\textbf{Signal Smoothing.}
To further improve interpretability and reduce measurement noise, a rolling mean digital filter with a window size of 86400 samples was applied (one full day of samples) after the conversion from voltage to VWC. While temperature contributed to the noise floor, it was not a large enough fluctuation to warrant a temperature correction algorithm. This is largely due to the sensors being completely buried whereas other sensors with a larger form factor, such as the commercial sensor studied in preliminary experimentation, would have experienced larger swings due to exposure to sunlight.

\subsubsection{\textbf{Sensor Longevity and Stability}} 
\label{sec:longevity} Galvanic soil sensors have been often dismissed in prior literature due to concerns about electrode corrosion and long-term signal drift under constant electrochemical load. To evaluate durability in our context, we monitored three deployed sensor nodes (Zn/SS) continuously for over 70 days, each measuring soil moisture every 20 minutes depicted in Figure \ref{fig:figure10}B, where dashed lines represent missing data due to indoor power loss. All sensors remained functional throughout the observation period, with no visible degradation in signal amplitude.
\vspace{0.1in}
\subsubsection{\textbf{Performance in Different Soil Types}} 
\label{sec:soiltypes}
Figure \ref{fig:figure9}C depicts sensor performance across varying soil types. Of significance are the varying drying rates of the soils. We tested 3 soil types courtesy of a local farm: Osco, a well-drained loess-derived silt loam, Catlin, a somewhat poorly drained silty clay loam, and Wyanet, a well-drained silt loam over sand and gravel.

This setup, depicted in Figure  \ref{fig:figure9}A,  was completed using 3 separated partitions of the different soil types inside a single container. Each partition received 2 zinc/stainless steel galvanic sensors, and was logged continuously over 37 days and 3 watering cycles. The figure depicts potting and Catlin soil types, both of which drain poorly, paralleling each other. Similarly, the well draining Osco and Wyanet soils also parallel each other. We additionally observed the test setup to be on a slightly uneven surface, encouraging some water to pool to one side, with each of the sensors on that side showing water draining more slowly.


\vspace{-0.1in}
\subsection{Power Availability}
\begin{figure}[!ht]
    \centering
    \vspace{-0.1in}
\includegraphics[width=\linewidth]{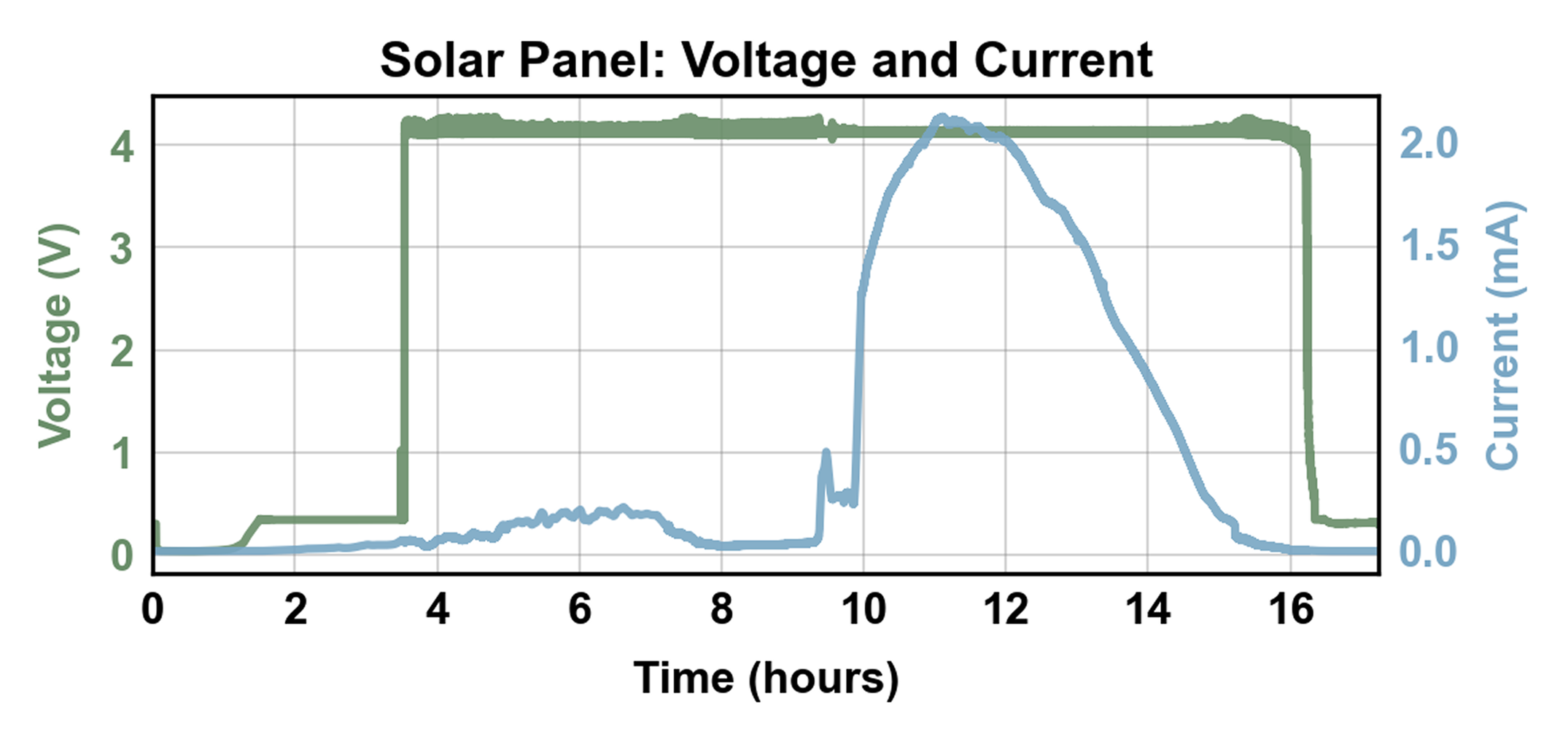}
    \vspace{-0.3in}
    \caption{Solar availability over one day, with an overcast morning and sunny afternoon. Note the easily identifiable cloudy morning with high voltage presence and no current which we use to identify a cloudy day. Later in the day, the panel shows a current of up to 2 mA as the sun hits the panel. This distinct behavior, especially the sharp rising voltage, is a product of the BQ25570 MPPT algorithm.}
    \label{fig:figure11}
\end{figure}

\begin{figure}[ht]
    \centering
 \includegraphics[width=\linewidth]{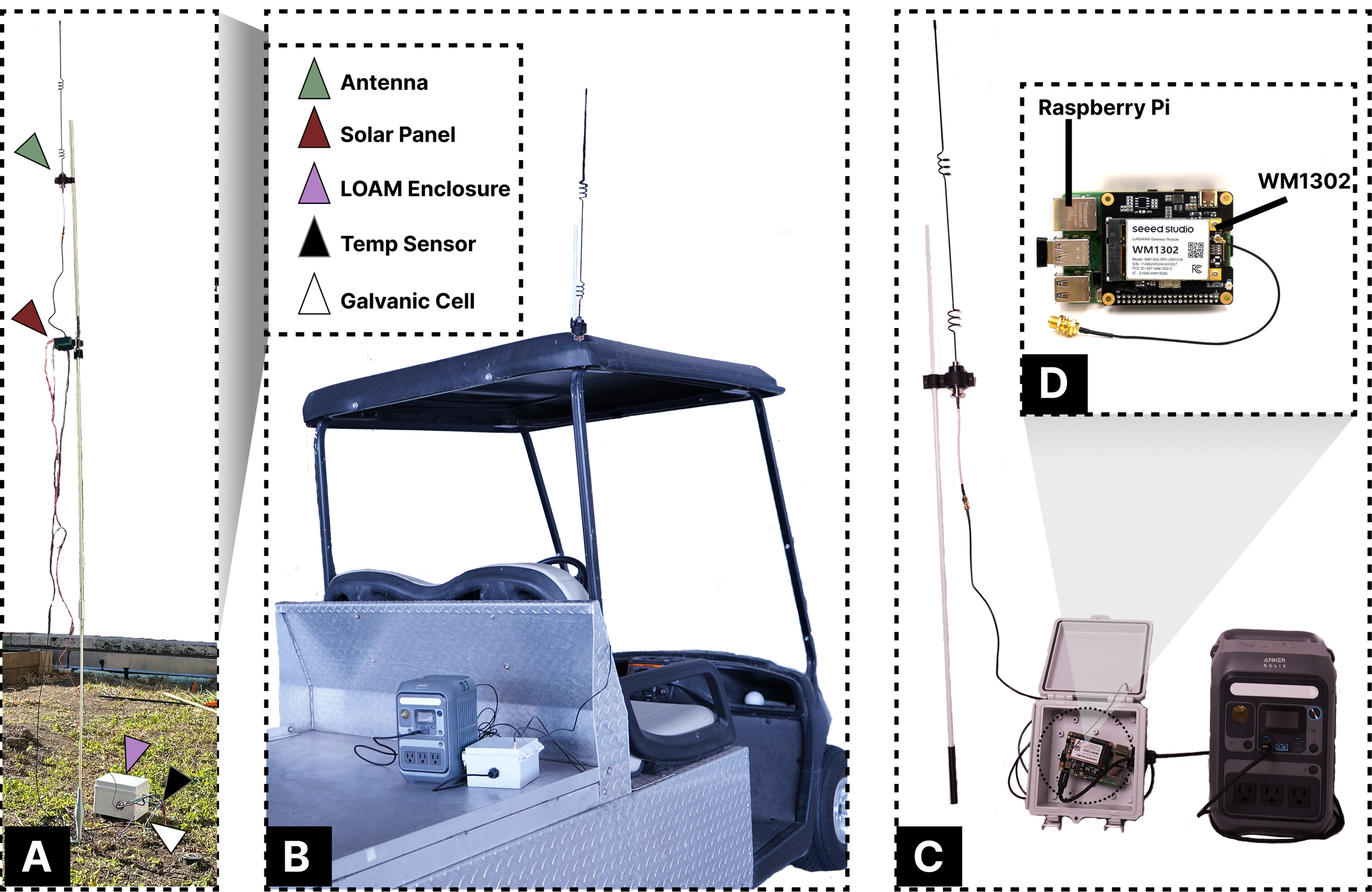}
    \caption{Scale-model outdoor deployment setup, with the sensor node (A) communicating directly across a courtyard from the mobile backend placed on the tailgate of a GATOR-like cart (B). Backgrounds removed for clarity given thin components. At right, a small enclosure (C) houses a lightweight Linux single-board computer with a Semtech WM1302 attachment (D).} 
    \label{fig:figure12}
\end{figure}

The system is designed to operate continuously under intermittent solar conditions using only a 1F supercapacitor as its energy store. We collected data on every possible cycle the FSM can follow to produce a detailed profile of the power consumption using a Joulescope precision power meter.  A full power breakdown—including instantaneous draw per subsystem and active duration—is shown in Table 1, and a solar harvesting profiles can be viewed in \Cref{fig:figure11}. The Microlink IMM solar panel  delivers enough current to restore full charge within 15 minutes of direct sunlight. Voltage-aware duty cycling and environmental context detection (via panel current and voltage) allow the FSM to skip transmission during low-power conditions, preserving data. 

\vspace{-0.1in}

\noindent\textbf{Power Consumption}
We present and discuss power consumption findings as a function of 20-minute cycles. Given the measured 15-minute charge-time of the capacitor , 20 minutes was selected as the next-highest cycle time evenly divisible into one hour, giving discussion numbers a realistic baseline in which the capacitor is afforded a forgiving time buffer. Further estimations using alternate length duty cycles is provided in Table 2. The fastest observe soil drainage cycle we measured occurred over 3 hours, leaving all of the presented estimations well under the minimum required sampling frequency. This intentional intermittency ensures deterministic behavior even during extended cloudy periods. The average power budget across a 20-minute cycle in which there is sunlight (and therefore the radio will transmit) is estimated at 360.47 uW (Active cycle A or C \Cref{tab:power_consumption}) , while the average power budget across a 20-minute cycle in which there is no sunlight (and therefore the MCU will shut down as quickly as possible) is estimated at just 8.14 uW. In understanding this, we sought to optimize functionality with the FSM to transmit only when actively charging. In doing so, measurements show that the node can sustain full sensing functionality for approximately 21 days in complete darkness.  

\noindent\textbf{Optimizing Power Consumption} To achieve reliable energy autonomy under constrained harvesting conditions, we conducted a systematic power profiling of all hardware components and system configurations. In particular, we examined how LoRa transmission parameters—including spreading factor (SF), coding rate (CR), transmission power, and payload size—affect the total energy cost of a cycle. Unsurprisingly, transmission power had the most dramatic impact on peak current draw, with higher dBm settings linearly increasing energy per bit.

Spreading factor significantly impacted airtime and energy use: increasing SF from 7 to 12 nearly tripled transmission duration, and thereby the active window of the MCU and SX1276 radio. While high SF values improve link robustness, we found that in low-noise, short-range conditions, SF7 with CR=4/5 provided the best tradeoff between delivery success and energy cost.
\vspace{-0.1in}
\begin{table}[!ht]
\centering
\small
\setlength{\tabcolsep}{1pt} 
\renewcommand{\arraystretch}{0.8}
\begin{tabular}{@{}p{0.6\columnwidth}r@{}}
\toprule
\textbf{Cycle Tracked} & \textbf{Cycle Consumption (\si{\milli\joule})} \\
\midrule
A: 1→1B→2→3→7 & 429.403 \\
B: 1→1B→2→3B→7 & 414.165 \\
C: 1→1B→2→3→3B→7 & 429.403 \\
D: 1→4→7 & 47.334 \\
E: 1→5→7 & 6.608 \\
F: 1→6→7 & 6.608 \\
\midrule
\textbf{Condition} & \textbf{Average Power Draw (\si{\micro\watt})} \\
\midrule
Active Cycles (A or C) & 360.47 \\
Inactive Cycles (E or F) & 8.14 \\
Ideal Sunny Day (18F, 1D, 34C, 1A, 18F) & 179.88 \\
\midrule
\textbf{Condition} & \textbf{Light Intensity (kLux)} \\
\midrule
Low Light & 4.5 and under \\
Cloudy/Diffuse & 5-12 \\
Bright & 30 \\
Full Sun & 80 \\
\bottomrule
\end{tabular}
\caption{A breakdown of the power consumed under each possible FSM cycle, using 20-minute cycles operating actively for 1 second and sleeping for 1199 seconds. Cycles, e.g., "18F" indicate how many consecutive cycles we estimated the FSM to direct, in this case 18 cycles of the 1→6→7 path. Also included are light intensity measurements under different environmental conditions.}
\label{tab:power_consumption}
\end{table}
\vspace{-0.25in}

\begin{table}[!ht]
\vspace{-0.2in}
\centering
\small
\begin{minipage}{\columnwidth} 
\setlength{\tabcolsep}{1pt} 
\renewcommand{\arraystretch}{0.7}
\begin{tabular}{@{}p{0.6\columnwidth}r@{}}
\toprule
\textbf{Duty Cycle (\si{\minute})} & \textbf{Capacitor Life (\si{\day})} \\
\midrule
20 & 21.81 \\
15 & 16.27 \\
10 & 10.75 \\
5 & 5.22 \\
1 & 0.80 \\
0.5 & 0.25 \\
\midrule
\textbf{Condition} & \textbf{9.68J Overhead Impact (\si{\joule})} \\
\midrule
1 Ideal Day (18F, 1D, 34C, 1A, 18F) & 0.128 \\
21 Consecutive Dark Days (1570F) & 9.546 \\
\bottomrule
\end{tabular}
\caption{Capacitor life estimates under different duty cycles and the impact on the capacitor energy overhead over a given time frame supposing 20-minute cycles.}
\end{minipage}
\label{tab:power_consumption_2}
\end{table}
\vspace{0.4in}

\noindent \textbf{Optimizing Duty Cycle} To identify how the choice of duty cycle impacts the operation of the node, we conducted capacitor life estimates to determine how long it would remain functioning under different duty cycles. These estimates were done assuming the same conditions described in \Cref{tab:power_consumption}, which is dark from 0000-0600, light from 0601-1800, and dark again from 1801-2359. These calculations revealed that, for more frequent duty cycles of 30 seconds, the lifetime of the node would be reduced to only a quarter of a day.

\subsection{\textbf{Exploratory Deployments}}
To verify the viability of the full sensing platform under real-world conditions, we performed a series of exploratory deployments after completing the core system stack. These trials were intended as as functional validation of the complete end-to-end system: nodes, protocol, gateway, backend, and data-handling logic.

\noindent \textbf{Corn Field Data Collection}
We visited a production corn site during peak canopy growth (mid-August) to collect soil samples from various locations of an active farm (soil map shown in \Cref{fig:figure8}b), enabling the lab-based data from  Osco, Wyanet, and Catlin soils discussed in \Cref{sec:soiltypes}. This farm visit also enabled the data collection for RF propagation. Experimentation proved 1km to be the maximum distance for a node with zero obstruction in the fresnel zone to successfully transmit at 2dBm. This distance dropped to $\approx 250\,\text{m}$ when the antenna was dropped to within the vegetation (\Cref{fig:figure2}. This successful experiment proved the need to optimize link quality with clear line of sight as well as power in order to achieve battery-free success. It also validated proprietary RF propagation planning tools \cite{cloudrf}, which use elevation and terrain data to help visualize link quality.

\noindent\textbf{Indoor Data Collection}
We deployed three sensor nodes indoors in 1-gallon soil pots, co-located with TEROS-12 reference probes. This environment provided controlled light exposure, consistent hydration schedules, and stable temperature conditions. The goal was to verify sensor signal fidelity and baseline node behavior under predictable power input. Additionally, we tested Osco, Wyanet, and Catlin samples to observe galvanic response and confirm that soil-dependent variation is handled correctly by the analog front end and digital processing stack.

\noindent  \textbf{Scale-Model Outdoor Deployment}
To validate FSM behavior and power margins, we deployed a set of nodes and a mobile gateway across our university campus. This small-scale model confirmed that nodes can independently harvest energy, store and forward measurements, and reliably complete handshake-based communication cycles when the gateway becomes available. These tests further demonstrated correctness of the communication protocol and local backend, confirming the feasibility of mobile gateway operation under realistic conditions.

 Across all settings, the system performed as designed — harvesting energy, sensing, executing intermittent state transitions, and reliably delivering data when communication became available. As a result, the remaining problem space centers on physical road-level route-finding and path and timing optimization rather than on core system correctness, and we leave this to future work.

\vspace{-0.1in}
\subsection{Potential Cost Evaluations}
\textbf{Affordability}
This device was produced for a reasonable \$33.68, with several bespoke components driving both costs and performance up (Table 3). The solar panels provide high power density for their size, but could reasonably be traded out for a slightly larger and less efficient but more economical panel. Similarly, most of our ICs were sourced through TI, whose performance generally matches the price, which can be further optimized.

\begin{table}[!ht]
\hspace{-10pt}
\centering
\small
\begin{minipage}{\columnwidth} 
\setlength{\tabcolsep}{1pt} 
\renewcommand{\arraystretch}{0.8}

\begin{tabular}{@{}p{0.77\columnwidth}r r@{}}
\toprule
\textbf{Category} & \textbf{Qty} & \textbf{Total (USD)} \\
\midrule

\rowcolor{gray!20}
\multicolumn{3}{@{}l}{\textbf{Node Items}} \\
Capacitors & 37 & 0.24 \\
Resistors & 16 & 0.12 \\
Inductors & 9 & 0.35 \\
Supercapacitor & 1 & 2.25 \\
ICs / Active Components & 12 & 13.86 \\
Connectors and Mechanical & 9 & 2.79 \\
Crystal / Timing Component & 1 & 0.43 \\
Antenna (+ Mast + Coax) & 1 & 8.76 \\
PCB Fabrication (10-board price) & 1 & 4.87 \\
\midrule

\rowcolor{gray!20}
\multicolumn{3}{@{}l}{\textbf{Gateway Items}} \\
WM1302 & 1 & 47.40 \\
Raspberry Pi & 1 & 35.00 \\
Antenna ( + Mast + Coax) & 1 & 8.76 \\
\midrule

\rowcolor{gray!20}
\multicolumn{3}{@{}l}{\textbf{Deployment Estimates}} \\
\multicolumn{2}{@{}l}{1 Node} & 33.68 \\
\multicolumn{2}{@{}l}{20 Nodes (100 Acre Plot)} & 673.56 \\
\multicolumn{2}{@{}l}{300 Nodes (15 × 100 Acre Plots)} & 10,103.37 \\
\multicolumn{2}{@{}l}{1 Basestation} & 91.16 \\
\bottomrule
\end{tabular}
\caption{Extended cost breakdown with deployment estimates using 5000-unit extended pricing.}
\vspace{-0.35in}
\end{minipage}
\label{tab:loam_extended_cost}
\end{table} 
\section{Related Work}
Our work sits at the intersection of three research areas, which together shape its design priorities of low cost, low power, a farm-friendly form factor and deployment.
\subsection{Soil Moisture Sensing }
Soil moisture sensors range from precise scientific instruments to single-shot probes, each balancing accuracy, durability, power, and cost. Scientific instruments quantify water content by measuring a physical property of water in soil, such as hydrogen atom density \cite{zreda_measuring_2008} or propagation velocity of electromagnetic signals (Time-Domain Reflectometry, or TDR) \cite{topp_electromagnetic_1980}. They provide accuracy and calibration flexibility but require high cost, power, and expert operation. Frequency-Domain Reflectometry (FDR) similarly infers soil moisture from the frequency response of a high frequency signal (around 100 MHz). Commercial sensors such as the TEROS-12 blend capacitive measurements with the Topp equation to map permittivity to water content \cite{METER_TEROS12_2018}, but draw tens of milliwatts, incompatible with battery free operation. In contrast, low-frequency capacitive sensors treat soil as a lumped capacitor in a simple RC or oscillator circuit and output an analog voltage correlated with volumetric water content. Electrical Conductivity probes are active, measuring conductivity instead of frequency response \cite{rhoades_electrical_1976}. Resistive sensors are cheaper \cite{bouyoucos_rapid_1926}, but suffer from poor accuracy and rapid electrode degradation. Galvanic sensors use dissimilar metal electrodes to generate an ion-dependent potential \cite{gaikwad_galvanic_2015}, yet drift from corrosion and changing ion activity. Our node targets these limitations with a high-impedance analog front end that stabilizes galvanic readout while meeting low power budget to enable battery-free operation. 
\begin{table*}[!t]
\renewcommand{\arraystretch}{1.5}
\centering
\small
\setlength{\tabcolsep}{6pt}
\begin{tabularx}{\textwidth}{
Y|Y|Y|Y|Y|Y|Y
}
\textbf{Work} &
\textbf{Sensor Type} &
\textbf{Networking} &
\textbf{Tx Power} &
\textbf{Device Power} &
\textbf{Power Source} &
\textbf{Duty Cycle} \\
\hline
\text{IoTree\cite{dang_iotree_2022}} & Tree Nutrients, Active  & Simplex LoRa & 20dBm & \SI{48.5}{\milli\watt} & Wind & None\\
Josephson et al.\cite{josephson_low-cost_2021}$^{2}$  & One-Shot Soil Moisture & Backscatter & N/A & \SI{342}{\micro\watt} & Battery & N/A \\
WAPPFRUIT\cite{barezzi_wappfruitautomatic_2024} & Soil Moisture, Capacitive  & Half Duplex LoRa & 10dBm & \SI{314}{\micro\watt} & Battery & 1hr \\
AeroEcho \cite{ren_aeroecho_nodate} & Platform Only & LoRa Backscatter & N/A & \SI{538}{\micro\watt} & Battery/RF & N/A \\
Loubet et al.\cite{loubet_lorawan_2019} & Concrete Health, Active & Half Duplex LoRa &  2dBm & \SI{100}{\milli\watt} & RF & N/A\\
Ding,  Chandra\cite{ding_chandra}$^{3}$ &  One-Shot Soil Moisture &  Wifi Backcatter & N/A & \SI{752}{\micro\watt} & Battery& N/A\\
Eclipse\cite{eclipse} & Air Quality, Active & Cellular & Not Reported & \SI{13.2}{\milli\watt} &Solar / Battery &1 min\\
\rowcolor{gray!15}
\textbf{This Work} & Soil Moisture, Temp., Galvanic & Half Duplex LoRa & 2dBm & \SI{180}{\micro\watt} & Solar & 20 min \\
\end{tabularx}
\vspace{0.5em}
\caption{Low Power Networked Sensors with full prototypes. $^{1}$ FloatingBlue's power was partially reported and done at an rx distance of several feet, estimation was required. $^{2}$ Josephson tested energy harvesting and opted for semi-passive for signal integrity. $^{3}$ Estimation was required assuming a 3.3V operating voltage with a capacity of 2000mAh over the one-year battery life reported.}
\label{tab:similar_works}
\vspace{-0.25in}
\end{table*}

\vspace{-0.15in}

\subsection{Agricultural Wireless Sensor Networks}
Wireless soil moisture sensing is a heavily studied area across the CPS, SIGMOBILE, and SIGCOMM communities. In practice, power, cost, and range jointly dictate which wireless method is feasible for deployment in agricultural fields. \textbf{Self-powered snapshot-based sensors}, such as powerless RFID tags are accurate, inexpensive, and scalable as long as a drone or human periodically visits and interrogates them, since their communication range is short. Prior work spans early RF-powered sensing \cite{benallegue_use_1995}, backscatter systems \cite{josephson_time--flight_2020,elkharrouba_surface_2022,daskalakis_uw_2018,yen_soil-powered_2023}, RFID-based tags \cite{pichorim_two_2018,wang_soil_2020,cappelli_battery-less_2021,hasan_towards_2015,kim_rfid-enabled_2014}, and radar-style approaches using mmWave \cite{zhuang_data-driven_2024,chen_soil_2022,chen_metasoil_2024} or UWB \cite{ding_soil_2023,malajner_soil_2019}. Next category is \textbf{ad hoc medium-range networks} (on the order of 10--100,m) built from commodity radios such as LTE \cite{feng_lte-based_2022}, BLE \cite{hanel_estimating_2024,keshavarz_dynamic_2025}, LoRa \cite{chang_sensor-free_2022,kiv_smol_2021}, and WiFi \cite{ding_towards_2019,nguyen_frequency-sensitive_2023,salman_wi-fi_2024}. These nodes are relatively low power, but their effectiveness depends on nearby infrastructure and a sufficiently dense mesh with gateways so that data can be forwarded and stored reliably. The third category is \textbf{low-power long-range networked sensors} (\Cref{tab:similar_works}), typically using LoRa \cite{chang_sensor-free_2022,kiv_smol_2021} or LTE \cite{feng_lte-based_2022,vasisht2017farmbeats}. These nodes trade higher power draw for scheduled transmissions, multiple onboard sensors, and extended range. However, they remain heavily dependent on permanent cellular or backhaul infrastructure and are often simplex \cite{dang_iotree_2022}, meaning no information is sent back to the node, and if a transmission fails, that data is permanently lost. 
In vegetated agricultural fields, even LoRa exhibits a strong power--range tradeoff, with practical ranges often limited to only a few hundred meters.
Killometer scale communication in dense vegetation at low infrastructure cost is still unsolved. 

\vspace{-0.15in}
\subsection{Opportunistic and Delay Tolerant Networks}
Opportunistic and delay-tolerant networking enable mobility-as- infrastructure, where designated carrier nodes ferry or ``mule'' \cite{shah_data_2003} data back to the internet. A common form is the \textbf{mesh network} \cite{noauthor_meshtastic_nodate}, where all nodes act as routers and compensate for limited range with peer-to-peer relaying \cite{chen_contact-aware_2020,ayele_towards_2018,florita_opportunistic_2020}. Mesh designs help cope with intermittent infrastructure but require nodes to stay powered , which is unrealistic for passive or ultra low power sensor networks.

\textbf{Delay Tolerant Sensor Networks (DTSN)} buffer data locally and transmit only when connectivity becomes available \cite{kietzmann_delay-tolerant_2022,baumgartner_loragent_2020}. They are used where reliable links are impractical, such as oceanographic and maritime monitoring, where buoys \cite{elgharbi_maritime_2025,tuyishimire_clustered_2019} or underwater vehicles \cite{desai_autonomous_2013,msaad_enabling_2021,fourniol_underwater_2023} hand off data to the first vessel they encounter, effectively mailing data back to shore. These systems operate under extreme latency rather than tight power budgets, yet show how rural agricultural WSNs might scale by leveraging mobility instead of new fixed infrastructure.

Mobile-first networking has also been explored for agriculture \cite{vasisht2017farmbeats,hossain_soil_2022,bhadauria_robotic_2011,paredes_lora_2023,sie_byon_2025}. These systems address non permanent infrastructure through mobile data collection but rarely support passive sensing, true low power communication, or the deeply duty cycled operation required for batteryless deployments. While data mules and delay tolerance are effective at the network layer, most DTSN designs still assume active, powered nodes and do not enable indefinite, maintenance free operation in energy harvesting sensor networks.
\vspace{-0.1in}

\section{Discussion}
\subsection{Limitations and Future Work}
\noindent\textbf{Power Budget and Parasitics.}
While our theoretical power budget projects an average draw of approximately 570 nA in deep sleep, measurements show actual consumption in the 800–900 nA range. This discrepancy likely originates from parasitics in the testing setup, surface leakages on the PCB, and software configurations at sleep-time. A single misconfigured pin can result in up to 500nA of leakage in sleep mode. Additionally, further optimization of capacitor type and material is required. Disclosure from the vendor indicates our selected capacitor self-discharges logarithmically, and would dominate power losses after a week without charging \cite{kemet}. Similarly, since the burying of the node renders form factor a non-issue, more capacitors could be employed with creative charging to minimize self-discharge. Finally, the dominant contributor of the 360.47 \si{\micro\watt} in active transmission mode is time spent in receive mode. Due to a LoRaWAN-standard 1-second delay for sending downlinks, our nodes spent between 90\% and 93\% of their active time at idle. Future work should explore usage of  emerging advances in BLE \cite{LR_BLE} to explore using our novel framework to bring the power savings of BLE to rural areas.

\noindent\textbf{Sensor Design and Analog Front End.}
Although the galvanic soil probe provided stable long-term readings, its dual sensitivity to pH and water content remains an open challenge. Rigorous experimentation with many nodes is required to create a robust model for VWC. Additionally, future designs should introduce  protect from electromagnetic interference protections. 

\noindent\noindent\textbf{Node Range and Data-Aware Path Planning.} The average USA farm is 500 acres, and farmers tend to several different fields at once, introducing further work into road-level route finding and path and timing optimization. The majority of this work's focus was on the Sensor and Connectivity Layers; future work exists in the Activity Layer, making an intelligent tool that utilizes farm plots, road maps, and node positions to automate the path planning process.

\noindent\textbf{Expanding to Other Measurements.}
Although our platform currently targets soil moisture, the same low-cost, activity-aware, battery-free platform can be extended to host sensors for pH, nutrients, salinity, and contaminants such as PFAS, enabling richer in-situ monitoring across large, infrastructure-poor farms.

\vspace{-0.1in}
\subsection{Emerging CPS/SenSys Research Themes}
\noindent \textbf{Materials-Centric Approach to System Design.}
Our work illustrates how translating advances in materials science into system design can directly optimize a key system parameter, in this case power. Materials scientists typically focus on improving the materials that constitute sensors, such as galvanic cells \cite{gaikwad_galvanic_2015}, while systems researchers tend to rely on off-the-shelf components and concentrate on computation and architecture. Our contribution bridges these two domains by rethinking the sensing stack end to end: we pair a galvanic sensor with a custom high-impedance analog front end, for the first time making this sensing modality suitable for ultra-low-power, field-deployable systems. 

\noindent \textbf{Human-in-the-Loop System Design.} Burden shifting is a familiar strategy in CPS/SenSys. Backscatter, for instance, moves the power-hungry carrier generation to infrastructure. Our platform advances this idea by making burden shifting explicitly human-in-the-loop: it leverages farmers’ routine movement to achieve installation and coverage instead of adding new infrastructure or labor. This points to a broader design space where systems are co-designed with everyday human activity to shift cost, power, and range.

\section{Conclusion}
This work contributes a new design pattern for battery-free, maintenance-free sensing at farm scale, where energy availability and human movement replace permanent infrastructure. By exploring mobility-as-networking, we have shown a viable alternative to high power RF transmission for data link quality. Future versions of this platform can push toward multi-year, unattended soil monitoring—an essential step toward sustainable data collection that truly disappears into the field. 
\bibliographystyle{ACM-Reference-Format}
\bibliography{references}
\end{document}